\documentclass{aa}  
\usepackage{float}
\usepackage{graphicx}
\usepackage{txfonts}
\usepackage{textcomp}
\usepackage[utf8]{inputenc}
\usepackage{rotating}
\usepackage{caption}
\usepackage{subcaption}
\usepackage{tikz}
\usepackage{adjustbox}
\usepackage{fancyhdr}
\begin{document} 
	\title{Hydrodynamical shear mixing in subsonic boundary layers and its role in the thermonuclear explosion of classical novae} 

    \author{Marco Bellomo
          \inst{1,2}
          \and
          Steven N. Shore
          \inst{2,3}
          \and
          Jordi Jos\'e
          \inst{4}
          }
          
	\institute{Institut f\"ur Theoretische Physik und Astrophysik, Christian-Albrechts-Universit\"at zu Kiel, Leibnizstr. 15, 24118 Kiel, Germany;\\	
    Dipartimento di Fisica ''Enrico Fermi" Universit\'a di Pisa, and INFN - Sezione di Pisa, largo B.  Pontecorvo 3, 56127, Pisa, Italy \\
	\email{mbellomo@astrophysik.uni-kiel.de.com}
		\and
    Dipartimento di Fisica ''Enrico Fermi" Universit\'a di Pisa, and INFN - Sezione di Pisa, largo B.  Pontecorvo 3, 56127, Pisa, Italy \\
	\email{steven.neil.shore@unipi.it}
 \and
 OATS - INAF, Via G.B. Tiepolo, 11
34143 Trieste, Italy
\and
	Departament de Física, EEBE, Universitat Politècnica de Catalunya, Av. Eduard Maristany 16, 08019 Barcelona, Spain;\\
	Institut d’Estudis Espacials de Catalunya, c/Esteve Terradas 1, E-08860 Castelldefels, Spain\\
	\email{jordi.jose@upc.edu}
    }
	
	\date{Received ---, ; accepted ---}
	
	
\abstract
	{The transition zone between the white dwarf (WD) envelope and a circumstellar accretion disk in classical novae, the boundary layer, is a region of strong dissipation and intense vorticity. In this strongly sheared layer, the hydrogen-rich accreted gas is expected to mix with the underlying WD outermost layers so the conditions for the onset of the thermonuclear runaway (TNR) in classical nova will be different from the the standard treatment of the onset and subsequent mixing.}
	{We applied the critical layer instability (CLI) to the boundary between a disk-accreted H/He zone and the C/O- or O/Ne - rich outer layers of a mass-accreting WD in a cataclysmic binary and then used the resulting structure as input to one-dimensional nuclear-hydrodynamic simulations of the nova outburst. }
	{We simulated the subsonic mixing process in two dimensions for conditions appropriate for the inner disk and a CO 0.8 M$_\odot$ and CO and ONe 1.25 M$_\odot$ WDs using the compressible hydrodynamics code PLUTO.  The resulting compositional profile was then imported into the one-dimensional nuclear-hydrodynamics code SHIVA to simulate the triggering and growth rate for the TNR and subsequent envelope ejection.}
    {We find that the deep shear driven mixing changes the triggering and development of the TNR.  In particular, the time to reach peak temperature is significantly shorter, and the ejected mass and maximum velocity of the ejecta substantially greater, than the current treatment. The  $^7$Li yield is reduced by about an order of magnitude relative to the current treatments.}
    {}	

    \titlerunning{Hydrodynamical shear mixing in novae}
    \maketitle


\section{Introduction}
Classical and recurrent novae are cataclysmic binary systems in which a white dwarf (WD, also hereafter referred to as the primary) accretes mass from a surrounding disk that is fed by a Roche lobe-filling companion. In the standard model (e.g., \citealt{bodeevans}, \citealt{Starrfield2016}, \citealt{Jordi2016}) the accumulated mass is increasingly compressed to the point where thermonuclear reactions are initiated under conditions of mild electron degeneracy. The reactions proceed through a thermonuclear runaway (TNR). The energy release heats the baryons (for the accreted gas, this is mainly H and He, while for the outer WD layers, it is either C and O or O and Ne enhanced) that develop strong thermal convection that cascades into turbulence and produces an upwelling of material.  

This sequence leads to explosive ejection of the accreted layers. The development of the TNR is preserved in the elemental abundance preserved in the ballistically expanding ejecta. 

The current treatment of the accretion assumes that the incoming gas settles quiescently onto the WD envelope. In this scenario, the triggering happens only at the base of the accreted layer through a sort of noise. In 3D, thermal convection initiates as local fingering or plumes that spread and drive secondary Kelvin-Helmholtz vortices and filamentation as the reaction rate increases and the larger scale buoyant vortices begin to mix (\citealt{Casanova2018} and reference therein).

The observed pattern indicates that substantial mixing occurs to enrich the accreted gas with matter from the WD envelope but how and when this occurs remains a debated aspect of simulations of the nova outburst.  The nucleosynthetic yield and luminosity of the TNR depend sensitively on the depth of this mixing which has been treated in a variety of ways in the literature.  A non-exhaustive list includes rotation coupled with the Kelvin-Helmholtz instability, ({\citealt{Belyaev2012}}), KHI from now on, thermal diffusion (\citealt{Townsend1958}, \citealt{Zahn1974}), or a thermal instability triggered by axisymmetric modes (\citealt{Goldreich1967}) or Eddington-Sweet circulation (\citealt{Endal1976}, \citealt{Pinsonneault1989}). \citet{PriaKo84} use diffusion as a source of mixing, while \citet{Starrfield1972} and \citet{Jordi1998} use pre-enrichment of the accreted material to mimic mixing with the underlying substrate.\footnote{This artificially increases the opacity of the accreted material and limits the amount of mass accreted in the envelope.}  In contrast, we explore this possibility by modeling TNR imposing a compositional profile for the initial conditions derived from a hydrodynamical simulation.

\section{Shear mixing mechanism(s)}
\citet{Kippenhahn1978} were among the first to exploit the transfer of angular momentum in the boundary layer with their ``accretion belt'', as a mixing agent. They discounted thermal gradient meridional circulation as too slow and acting against the molecular weight gradient.\footnote{The treatment is essentially axisymmetric and lacks meridional flows. The disk and boundary layer are, therefore, treated as two-dimensional structures in which some turbulent motions are possible.} This was extended by \citet{Kutter1989} using similar arguments to obtain the spin-up mass fraction in the WD's envelope. 

\citet{Alexakis2004a} examined, instead, the so-called critical layer instability (hereafter CLI) as the mixing agent in the boundary layer. They proposed that convection within the accreted stratum generates gravity waves that, breaking, mix the WD envelope with the gas at the inner portions of the accretion disk. Consequently, the mixing in \citet{Alexakis2004a} is driven by buoyancy and sets in only when the layer turns convective. 

The CLI derives from the resonant interaction between superimposed flows when the maximum streamwise velocity is the same as a gravity wave generated at the interface \footnote{\citet{Glatzel1988}, \citet{Rosner2001}, \citet{Alexakis2004a} and \citet{Belyaev2012} assume a velocity discontinuity between a supersonic layer and the underlying WD, but this may only be applied to the initial conditions when the accretion disk is established; one would expect that the growing instability smooths this out.}.

This contrasts with the KHI which has no characteristic speed that acts around the velocity discontinuity, shearing the layers. Vortices grow independent of the gravitational acceleration and the center of vorticity remains stationary while the amplitude increases. Inertial effects from density contrast between the layers break the symmetry of the dynamics by the inertial term in the equations of motion but without resonances. In contrast, in the CLI the boundary ascends until the speed of the shear flow at the critical layer matches the speed of the gravity wave $v_{grav}$ generated at the interface since the gravitational acceleration $g$ provides the restoring force and $v_{grav} \sim (g \delta z)^{1/2}$ for a displacement $\delta z$. 
The shear stress at the boundary is proportional to the vertical gradient of the horizontal velocity and this generates a pump, well-known from the Ekman effect (e.g. \citealt{Rieutord1992}, \citealt{RieutordZahn1995}).

The driving depends on the slope of the perturbation, in contrast to the KHI where the growth is driven by the lift that peaks at the apex of the perturbation (\citealt{Alexakis2004a}), so there is a $\pi/2$ phase shift between the two instabilities. The growing vorticity lifts that layer and excavates the lower strata, mixing the material into the upper parts of the boundary layer. The acceleration is analogous to the Magnus force in vortical flows, $\rho {\mathbf\omega} \times {\bf v}$, where ${\bf \omega}$ is the shear (vorticity) for a velocity field ${\bf v}$  (see, for instance, \citealt{Lighthill1962}). The vorticity can be approximated with the sheared flow above the interface, ${\mathbf \omega} \sim [d^2{\bf v}(z)/dz^2]h$ (where ${\bf v}$ is the freely streaming, vertically sheared superimposed boundary wind that depends only on the vertical coordinate $z$, and $h$ is the height of the critical layer). Thus, the rate of energy transfer from the disk to the envelope is proportional to the wind shear and the horizontal wavelength $L$. 
This is essentially different from the KHI where the lift is provided by the maximum of the shear and the peak acceleration is at the crest of the vortex. In the CLI, it is at the point of maximum curvature of the wave so while the growth rate for the KHI depends on $dv/dz$, that of the CLI  depends on $d^2v/dz^2$. 

Convective mass transport depends on the temperature gradient in the burning zone. Its growth is on a buoyancy timescale.  In contrast, if there is dynamical pre-mixing, the triggering occurs in an already fuel-enriched environment that will develop on a much shorter thermal conduction timescale. This induces more rapid convection because the initial temperature gradient is higher so it is not unreasonable to consider that a pre-mixed layer would trigger and burn differently (\citealt{Alexakis2004b}). The physical process is much like the action of a carburetor in an internal combustion engine, in which the fuel is pre-mixed air to increase the explosive yield when ignites (e.g., \citealt{Lumley2001}). 

The essential difference between the two mechanisms is buoyancy. For two shearing layers with different densities, the center of vorticity will be displaced to the center of mass for the KHI. There is no characteristic speed or length scale. The CLI, instead, differs because gravity acts as a restoring force and controls the vertical density stratification. A resonance is reached at a height above the interface, where the velocity that produces the shear in the driving layer is the same as a gravity wave supported by the local hydrostatic structure. 

\section{Details of the simulations}

We performed the explosion simulations in two steps.  The first used PLUTO to obtain the shear-induced chemical profile, based on the WD envelope composition. The second imposed this profile on the one-dimensional, spherical SHIVA grid to continue the evolution through the onset of the TNR.

\subsection{Step 1: hydrodynamics from PLUTO}

\begin{figure}[h]
\centering
	\includegraphics[width=\hsize]{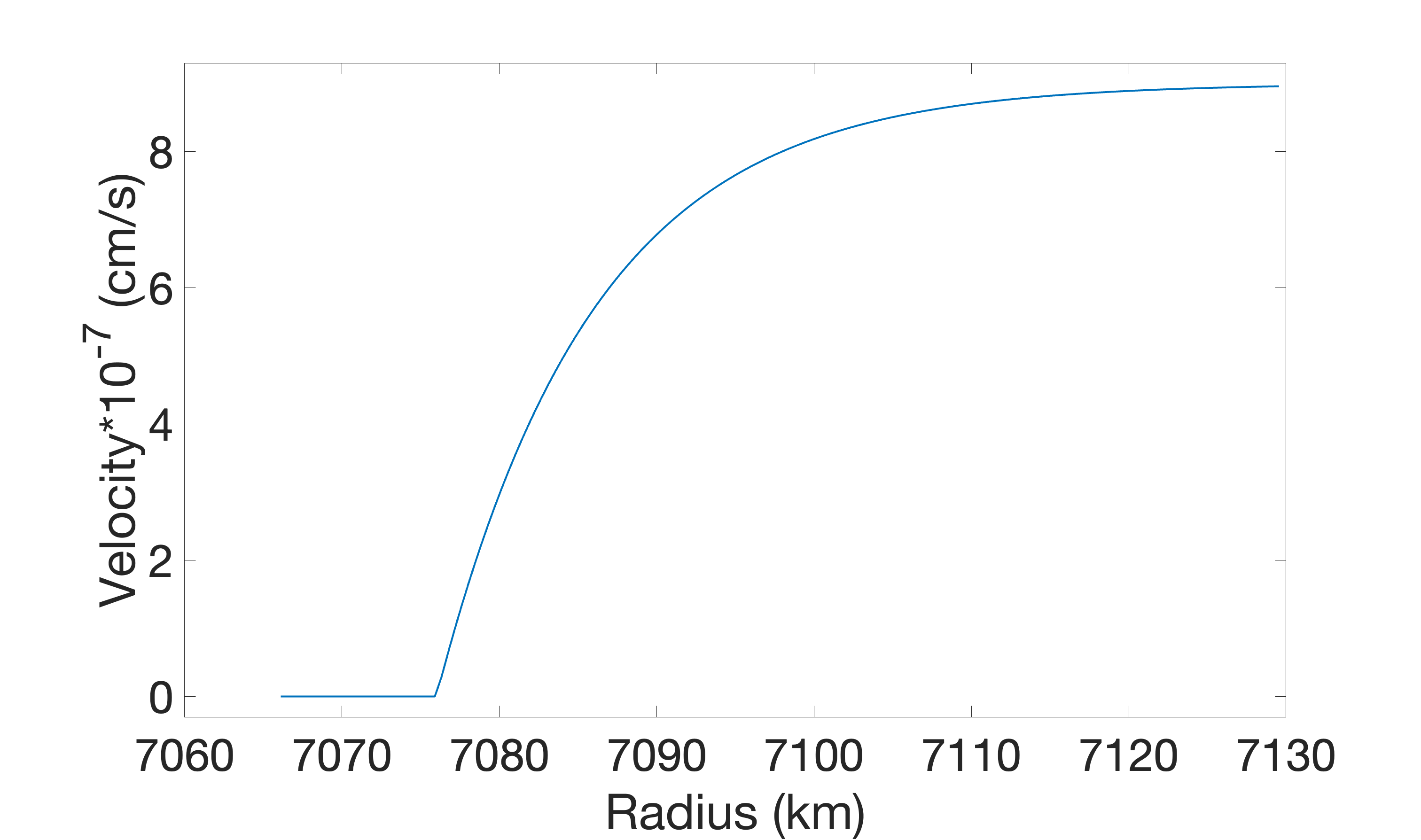}
    \caption{Velocity profile described from the eq. \ref{velocityprofile} for PL1.}
    \label{fig:velprofile}	
\end{figure}

We used PLUTO (\citealt{Mignone2007}), an open-source finite volume, compressible fluid dynamics code, to simulate the development of the CLI through saturation and breaking. We adopted the same initial conditions as \citet{Alexakis2004a} and imposed periodic horizontal and vertical reflecting boundary conditions for our 2D computational domain.
We selected the HLLC Riemann Solver assuming a three-wave model, and a second-order TVD Runge-Kutta time-marching scheme to advance the solution.  
We began with two superimposed compressible adiabatic (barotropic) fluids of different
density (composition), stratified according to hydrostatic equilibrium with:
\begin{equation}
	\rho=\rho_{i}\Bigl[1-(\gamma-1)\frac{g\rho_{i}y}{P_{0}\gamma}\Bigr]^{\frac{1}{\gamma-1}}
\end{equation}
\begin{equation}
	P=P_{0}\Bigl[1-(\gamma-1)\frac{g\rho_{i}y}{P_{0}\gamma}\Bigr]^{\frac{\gamma}{\gamma-1}}
\end{equation}
where $\rho_{i}$ and $P_{0}$ are the density and the pressure immediately above or below the interface, and $\gamma$ is the ratio of specific heats. In all our simulations we started with a velocity profile for $U = U(y)\hat{x}$ of the form:
\begin{equation}
    \label{velocityprofile}
	U(y)=\left\{
		\begin{tabular}{cc}
			$U_{max}(1-e^{-(y-L/2)/\delta})$ & $y>L/2$\\
			$0$						   & $y<L/2$
		\end{tabular}
	\right.
\end{equation}
where $L$ is the $y$ dimension of the computational box, and $\delta$ is a characteristic length scale, left as a free parameter. For computational convenience we placed the interface between the two layers at $L/2$. The two fluids are the C/O-rich outer white dwarf layer and the H/He accreted layer. The two fluids have a density ratio  $r=\rho_{1}/\rho_{2}$ where the subscript 1 stands for the upper fluid, and 2 stands for the lower. We used  $\rho_{1} = 360$ $g/cm^3$ and $\rho_{1} = 340$ $g/cm^3$, with $r = 0.6$ in the simulations for the initial structure, at the onset of accretion, as computed with the SHIVA code. 
The form of the forced shear profile, Fig. \ref{fig:velprofile}, is chosen for the boundary layer following studies of the wind-wave resonant instability in oceanography (e.g. \citealt{Miles1957}). Instead of a discontinuity in the transverse velocity, the  shear layer has a width of about one hundred pixels so the CLI also develops along with the initially excited KH modes.
The models, designated PL1 and PL2, are PLUTO simulations for  a 0.8 $M_{\odot}$ and 1.25 $M_{\odot}$, respectively. The model parameters are listed in Table\ref{setup}. 

We seeded the instability with a random perturbation across the surface with an amplitude of order $10^{-7}$ of the horizontal velocity to replicate the convective initial conditions. In contrast, \citet{Alexakis2004a} initialized the perturbation at the most unstable modes. A consequence of the reflecting vertical boundaries with no inflow at the upper boundary is an artificial initial upward displacement of the interface. There is an energy influx from the shear flow to the generated gravity waves but these transients rapidly damp.

\begin{figure}[h]
\centering
	\includegraphics[width=1.2\hsize]{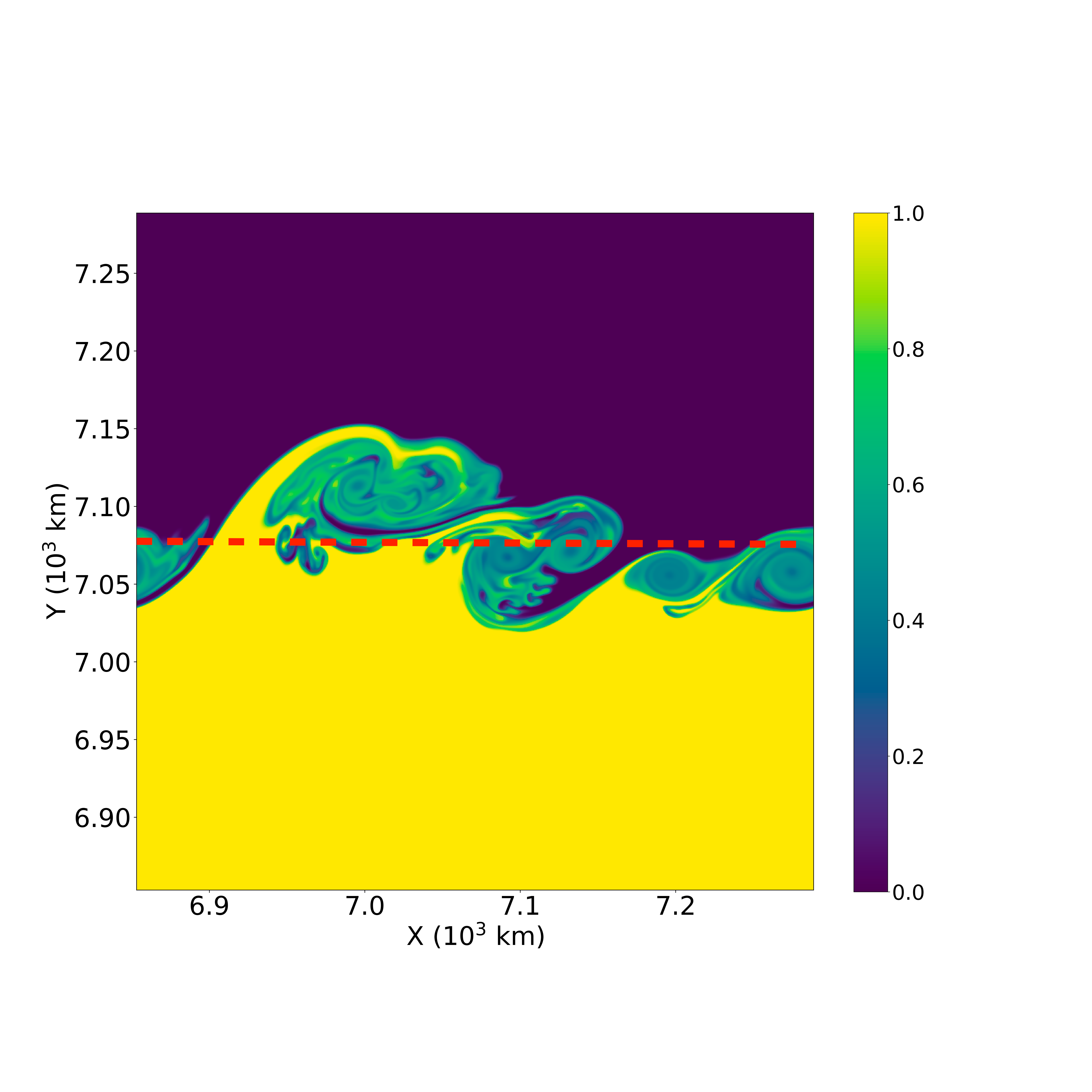}
\caption{$^{12}C$ mass fraction in simulation PL1 around discontinuity surface.}
\label{fig:trvort}	
\end{figure}

Our treatment of the boundary layer is two-dimensional. This is a compromise. In the absence of a fully thermodynamical simulation for the accretion with PLUTO (the simulation of accretion starts later with SHIVA), we chose to impose a shear profile as an initial condition and relax the flow as a thin disk. This is, in effect, integrating through a scale height so our densities are actually equivalent to column densities. The lower dimensionality induces an inverse cascade in which smaller vortices merge to the larger scale but the vorticity production cascades directly (e.g. \citealt{Krai1980}). The zonal integration of the tracer\footnote{The passive tracer $Y$ in PLUTO obeys a source-free continuity equation 
\[
\frac{\partial(\rho Y)}{\partial t}+\nabla\cdot(\rho Y\mathbf{v})=0
\]
that, in our case, is the fractional enhancement of $^{12}$C: +1 when we have only $^{12}$C and 0 when we have only accreted gas.} with depth produces the same compositional gradient as the fully turbulent precursor stage (Figs. \ref{fig:trvort}-\ref{fig:rhoev}). 

\begin{table}
	$$ 
	\begin{array}{cccccc}
		\hline
		\noalign{\smallskip}
		\text{Sim} &  U_{max}(cm/s)   & \delta(cm)     & M (M_{\odot})        & R (cm) \\
		\noalign{\smallskip}
		\hline
		\noalign{\smallskip}
		PL1   & 2\cdot 10^{8} & 1 \cdot 10^{4}    & 0.8 M_{\odot} & 7.056\cdot 10^{8}   \\
		PL2   & 2\cdot 10^{8} & 1 \cdot 10^{5}    & 1.25 M_{\odot} & 3.782\cdot 10^{8} \\
		\noalign{\smallskip}
		\hline
	\end{array}
	$$ 
	\caption{PLUTO simulations set up}
	\label{setup}
\end{table}

Although we cannot dynamically continue this with PLUTO, we can export this profile for the tracer into the envelope structure imposed on the WD at the start of accretion in the one-dimensional SHIVA code. We have the example in Fig. \ref{fig:rhoev} for the PL1 and PL2. Imposing the mixing profile at the start of the accretion phase ensures that the layer maintains a composition gradient in the run-up to the onset of convection and nuclear burning although, since the models are one-dimensional, no dynamic interaction can be included between the induced convection and the boundary layer. For completeness, we made a comparison with one explosion sequence for the 1.25 M$_\odot$ model, where the convection is treated using the ``12321'' prescription (\citealt{Jose2020}). This approach models convection based on time-dependent convective velocities and masses dredged up from the underlying white dwarf, extracted from three-dimensional simulations.

\subsection{Step 2: evolution from SHIVA}
\begin{figure}[!ht]
	\centering
	\includegraphics[width=0.99 \hsize]{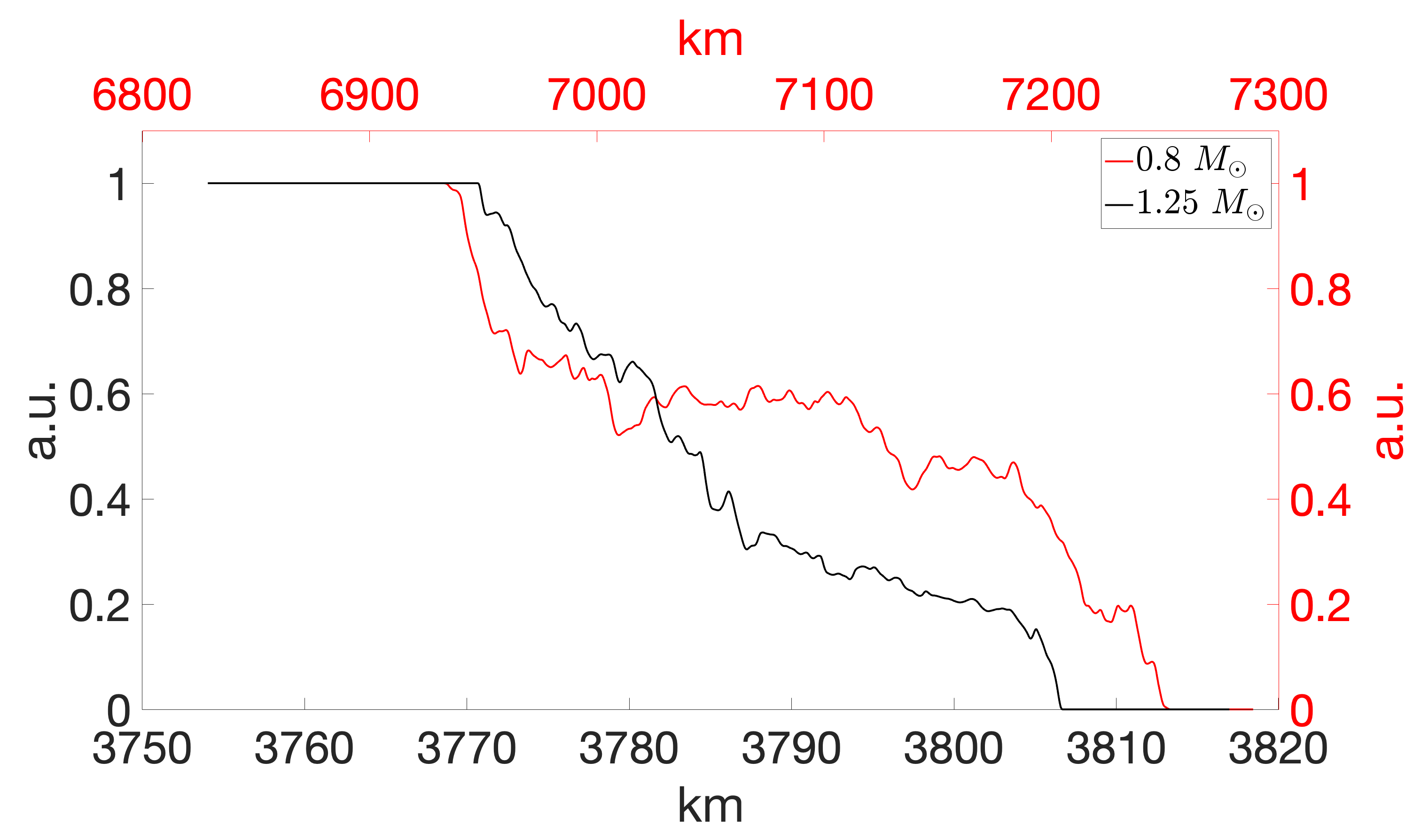}
	\caption{$^{12}C$ mass fraction as a function of height for different times, for PL1 and PL2.}
	\label{fig:rhoev}
\end{figure}

\begin{figure}
\begin{tabular}{@{}cc@{}}
    \subfloat[]{
        \label{}%
        \includegraphics[width=0.77\textwidth]{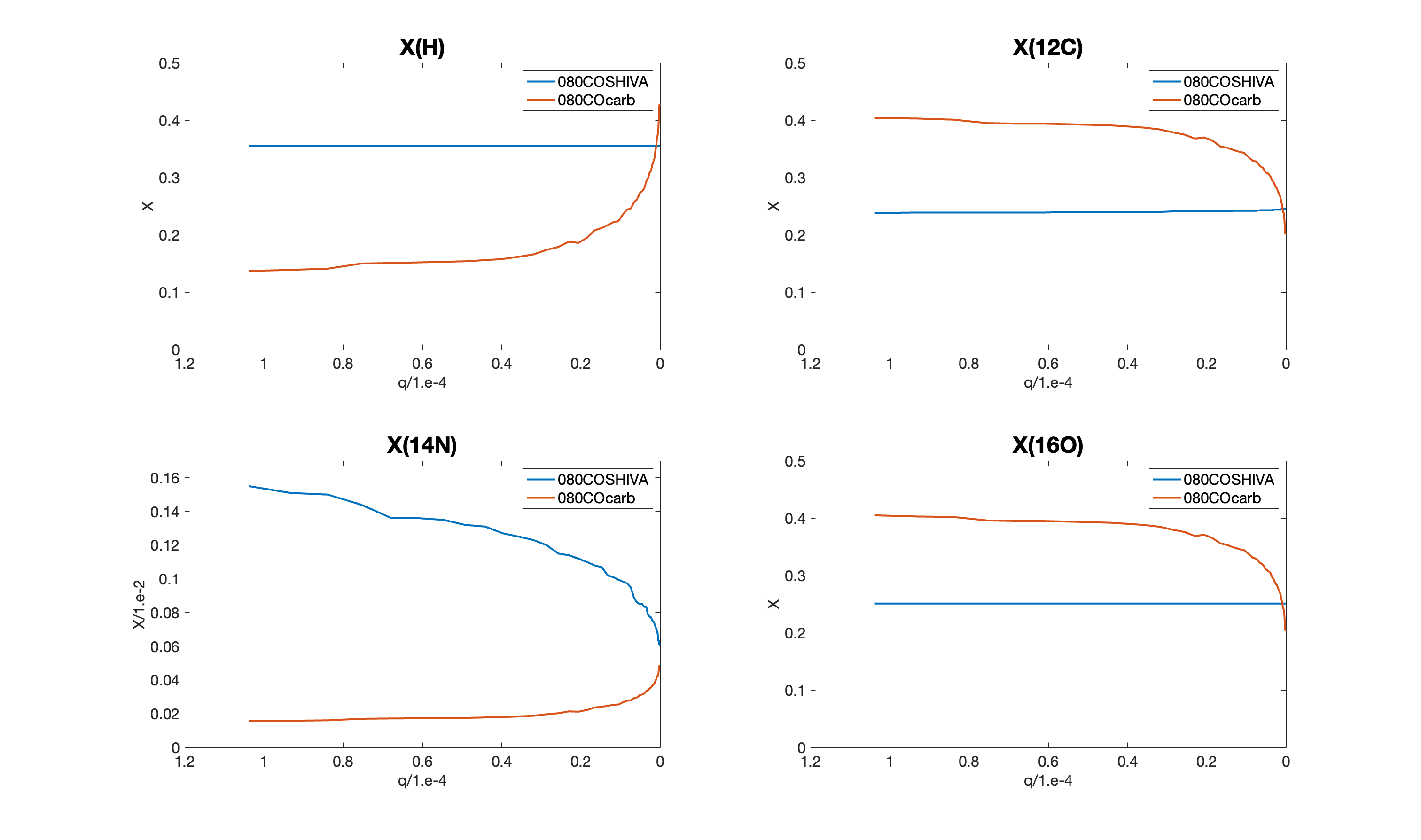}%
    }\\
    \subfloat[]{
        \label{}%
        \includegraphics[width=0.77\textwidth]{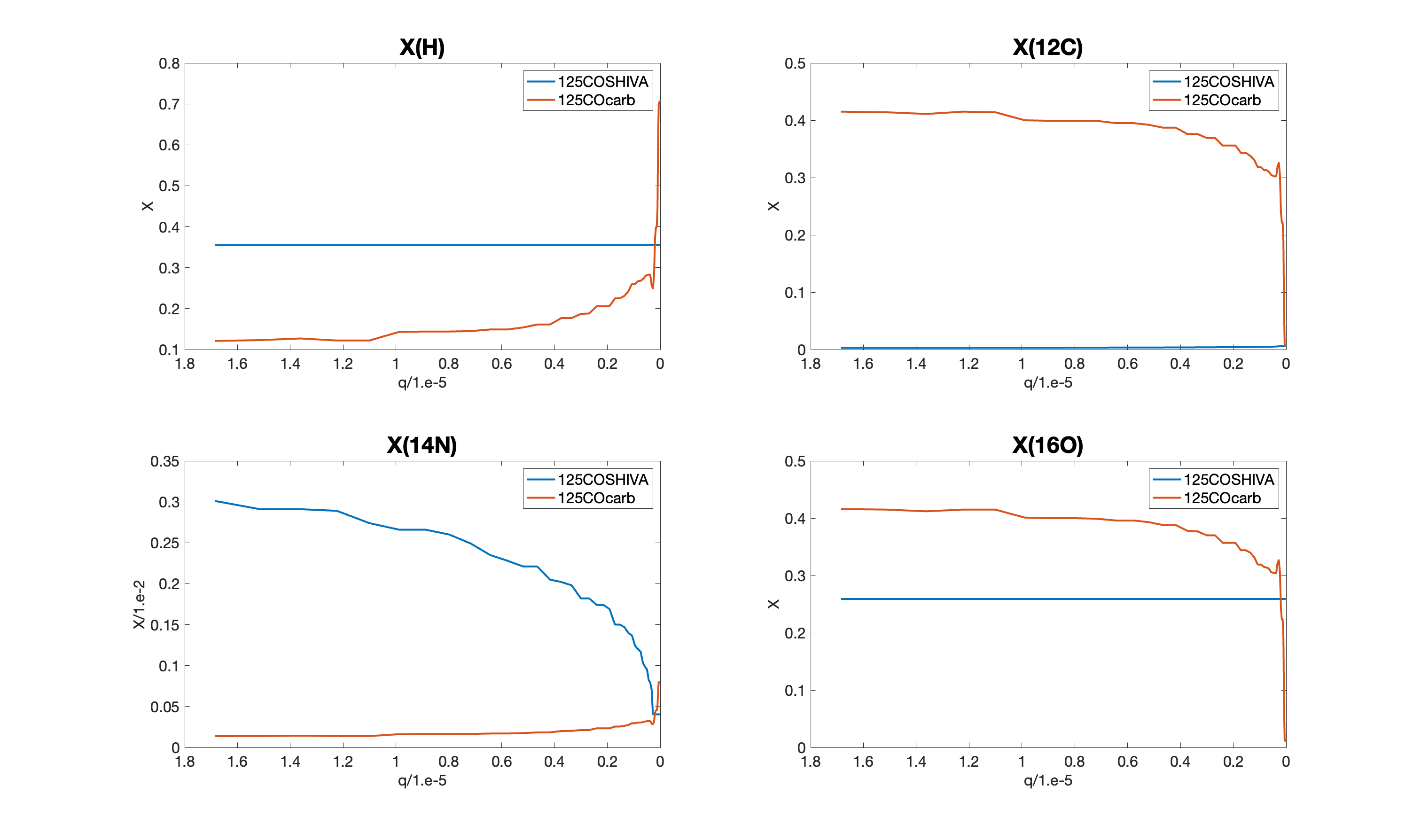}%
    }
\end{tabular}
\begin{minipage}{0.2\textwidth}
\caption{Panel (a): Initial mass fractions of $^1H$, $^{12}C$, $^{14}N$ and $^{16}O$ at the WD envelope for models 080COSHIVA, 080COcarb. The variable $q$ is a mass variable defined as $q = 1 -  M_{int}/M_{WD}$, where $M_{int}$ is the interior mass ($q=0$ denotes the outermost envelope layers - i.e. WD surface). Panel (b): same of panel a for models 125COSHIVA and 125COcarb.}
\label{fig:imrcomp}	
\end{minipage}
\end{figure}

We used the passive tracer in PLUTO (see Fig. \ref{fig:rhoev}) to calculate the final $^{12}$C mass fraction across the WD envelope, and determined an abundance-weighted, zonal averaged compositional mixing profile (see Fig. \ref{fig:imrcomp}). This was exported to SHIVA, an implicit 1D (spherically-symmetric) Lagrangian nuclear-hydrodynamic code, extensively used for the modeling of nova explosions (\citealt{Jordi1998}; \citealt{Jordi2016}).\footnote{For a description of a parallelized version of the SHIVA code see \citealt{Martin2018}.} The code solves the standard set of differential equations of stellar evolution, i.e., conservation of mass, momentum, and energy, and energy transport by radiation and (time-dependent) convection. The equation of state includes contributions from the electron gas with different degrees of degeneracy, a multicomponent ion plasma, and radiation. Coulomb corrections to the electron pressure are taken into account. Radiative and conductive opacities are included. For the simulations reported in this paper, a nuclear reaction network was used that contains 118 nuclear species, ranging from hydrogen ($^1$H) to titanium ($^{48}$Ti), linked through 630 nuclear interactions (mostly in the form of proton captures and beta disintegrations), with STARLIB rates  (\citealt{Sallaska2013}). Screening factors were also included. The final limitation of the present calculations is the absence of accretion heating but this is the assumption generally adopted in the one-dimensional models.

There are two broad flavors of classical novae, CO or ONe, distinguished by the WD composition.  Consequently, although material accreted from the nondegenerate companion has approximately solar composition (with a metallicity $Z \sim 0.02$),  the processing during the outburst stage is different (\citealt{Jose2020}).  
The SHIVA simulations were performed for two stellar masses, 0.8 and 1.25 M$_\odot$ (see Table \ref{SHIVA}), representative of CO and ONe white dwarfs, respectively.  The WD was initially relaxed to guarantee fully hydrostatic equilibrium conditions. The initial white dwarf luminosity was set at $10^{-2} L_\odot$. In models 080COSHIVA and 125ONeSHIVA solar composition  (\citealt{Lodders2009}) was assumed for the gas transferred from the companion at a constant rate of  $2 \times 10^{-10}$ M$_\odot$ yr$^{-1}$ and was assumed to mix with material from the outer layers of the underlying white dwarf to a uniform mixing ratio of 50\% (CO-rich material, in model 080COSHIVA; ONe-rich material, in model 125ONeSHIVA). For completeness, two additional models were computed: (i) Model 125COSHIVA, identical to Model 125ONeSHIVA, but assuming a CO WD, and (ii) Model 125COSHIVA123, in which the accreted layer onto an ONe WD is assumed to be purely solar, but once the entire envelope becomes fully convective, dredge-up driven by Kelvin-Helmholtz hydrodynamic instabilities sets via the ``12321'' scheme, increasing the metallicity of the envelope (see e.g., \citealt{Casanova2011}, \citealt{Jose2020}).

Finally, Models 080COcarb and 125COcarb were evolved adopting the chemical profile from step 1 (i.e., PL1 and PL2) with the structure relaxed by SHIVA to guarantee initial hydrostatic equilibrium with the new composition. The configuration was subsequently evolved with  SHIVA through the explosion, expansion, and ejection stages. The PLUTO-generated chemical profiles were assumed to be self-similar, and were maintained fixed during the overall accretion phase by resetting the composition of the envelope at every time step.

\begin{table}
	$$ 
	\begin{array}{ccc}
		\hline
		\noalign{\smallskip}
		\text{Model} & M (M_{\odot}) & \text{Composition} \\
		\noalign{\smallskip}
		\hline
		\noalign{\smallskip}
		\text{080COSHIVA}  & 0.8 & \text{50\% solar, 50\% CO (pre-en.)}   \\
		\text{080COcarb}   & 0.8   & \text{PL1 (shear)} \\
        \text{125ONeSHIVA}   & 1.25  & \text{50\% solar, 50\% ONe (pre-en)}\\
        \text{125COSHIVA}   & 1.25  & \text{50\% solar, 50\% CO (pre-en)}\\
        \text{125COcarb}   & 1.25  & \text{PL2 (shear)}\\
        \text{125COSHIVA123}   & 1.25  & \text{solar + dredge-up (12321)}\\
		\noalign{\smallskip}
		\hline
	\end{array}
	$$ 
	\caption{SHIVA simulations}
	\label{SHIVA}
\end{table}

\section{Discussion} 
In multidimensional simulations, the core-envelope interface becomes convectively unstable once burning initiates (e.g., \citealt{Casanova2018}). This dredges up CO-rich WD matter that pollutes the overlying accreted layer as the WD envelope is also supplied with H-rich gas but this counter-gradient downward mixing is weaker because the driving is only buoyancy. In contrast,  shear generates penetrative turbulence that is not as strongly impeded by the temperature gradient and the resulting excavation profile extends below the initial hydrostatic interface and the hydrogen is more intensely and deeply mixed.

 Our simulations address how the initial density discontinuity between the two layers is modified to a profile similar to the one in Fig. \ref{fig:rhoev} but the precise form of the final stage is, at least partly, an effect of the boundary conditions. The PLUTO simulations show two different stages of mixing. The first is dominated by small-scale eddy generation, and the second by turbulent diffusion. The mixing is quantified by averaging the density over the initial upper volume. Our results are qualitatively similar to \citet{Alexakis2004c}, who used FLASH, which is also an Eulerian code.  The mixed fraction grows nearly linearly in the dynamical range and then saturates when filamentation becomes pronounced. Once fully developed turbulence appears, the eddies dominate the viscosity.

In the usual treatment, the accreted layer is homogenized to a fixed fraction coming from the WD envelope. The shear mixing is a dynamic process that also transports hydrogen into the envelope, that is at higher pressure. Thus, while the mean is about the same, the effect of the greater enrichment of fuel in both the upper and lower strata because of the dredging relative to the homogenized layer, produces quantitatively and qualitatively different effects. These are listed in Tabs \ref{outputSHIVA}-\ref{outputShiva2} and shown in Figs. \ref{convcomp}-\ref{Enuclcomp}.

\subsection{Implications for nuclear yields}


For nova conditions, $^7$Li synthesis is expected to proceed through  beryllium formation (\citealt{Cameron1955}), initiated by $^3$He($\alpha$, $\gamma$)$^7$Be, which decays into the first excited state of $^7$Li by electron capture ($t_{1/2} \sim$ 53 days). The amount of $^3$He left over after the initial stages of the explosion, and the characteristic timescale to achieve peak temperature are critical for $^7$Be, and $^7$Li production. At $T > 10^8$ $K$, the $^8$B($\gamma$,p)$^7$Be photodisintegration reaction dominates over $^7$Be(p, $\gamma$)$^8$B  (\citealt{Hernanz1996}). Therefore, a rapid rise  to temperatures of about $10^8$ $K$ favors survival of $^7$Be, since exposure of $^7$Be to destructive (p, $\gamma$) reactions is reduced.  This is the case, for instance, for envelopes highly enriched in $^{12}$C since the most important reaction that triggers the TNR is  $^{12}$C(p, $\gamma$) (\citealt{Josè2016}). As shown in Table \ref{Litium}, for CO nova models, the more massive white dwarf (1.25 $M_{\odot}$ vs. 0.8 $M_{\odot}$) produces more $^{7}$Li (by a factor of 10 with the standard model and a factor of 100 with the carburetor model) because of the shorter time to reach peak temperature. In contrast, models of ONe novae, since they are not so enriched in $^{12}$C, have lower $^7$Li yields (e.g., 125ONeSHIVA vs. 125COSHIVA, both for 1.25 $M_{\odot}$).  All of this is with the standard SHIVA conditions at the triggering of the  TNR.

Implementing the new models presented in this paper we get both a faster rise in temperature during the TNR and a lower initial $^3$He content, that produce less $^7$Li for the low mass model, about one percent of the standard model. Indeed the lower initial $^3$He content overcomes the fast rise to peak temperature, leading to smaller $^7$Li production than the standard models with pre-enrichment. For the $1.25$ $M_{\odot}$ in the CO models, the pre-mixed case also yields about one percent that of the standard model, but almost a factor of 10 more compared to models with mixing driven by 3D Kelvin-Helmholtz secondary hydrodynamic instabilities (\citealt{Jose2020}).  The enrichment generated by mixing in the carburetor models is more than the temperature.

\onecolumn
\twocolumn

\onecolumn
\begin{table}
	$$ 
	\begin{array}{ccc}
		\hline
		\noalign{\smallskip}
		\text{}   & \text{080COSHIVA}    &   \text{080COcarb} \\
		\noalign{\smallskip}
		\hline
		\noalign{\smallskip}
		t (40\rightarrow 100 MK) [s]   & 6.72\cdot 10^4 & 2.5079\cdot 10^4\\
		t (100 MK\rightarrow T_{peak}) [s]   & 189 &  43.2 \\
        T_{peak}\text{ }[10^8 K]   &   1.49    &    1.80\\
		T_{base} \text{ fully convective envelope } [10^7 K]   & 8.39   &8.26\\
        M_{ejec}[10^{-5}M_{\odot}]   &   6.71  & 9.34 \\
        X(H)_{ejec}    &  0.331     &    0.14\\
        X(12C)_{ejec}  &   6.07\cdot 10^{-2}   &0.48\\
        X(13C)_{ejec}  &   0.118       &   0.30\\
        X(14N)_{ejec}  &   9.41\cdot 10^{-2}    &   7\cdot 10^{-2}\\
        X(15N)_{ejec}  &   1.01\cdot 10^{-3}    &   1.3\cdot 10^{-3}\\
        V_{ejec} (\text{mass-averaged}) [km/s]    &   1218    &   2244\\
        \text{Max } V_{ejec} \text{ when } R=10^{12} cm [km/s] &   2548  &3648\\
        K_{ejec} [10^{45} ergs]   &   1.23    &    5.53\\
        t (T_{peak}\text{ to }R=10^{10} cm)[s]  &   1077    &   412\\
		\noalign{\smallskip}
		\hline
	\end{array}
	$$ 
	\caption{Results for the $0.8$ $M_{\odot}$ models}
	\label{outputSHIVA}
\end{table}
\begin{table}
	$$ 
	\begin{array}{ccccc}
		\hline
		\noalign{\smallskip}
		\text{}   & \text{125ONeSHIVA}   & \text{125COSHIVA}  &  \text{125COcarb}\\
		\noalign{\smallskip}
		\hline
		\noalign{\smallskip}
		t (40\rightarrow 100 MK) [s]   & 7.03\cdot 10^{5} & 10499 &   7080\\
		t (100 MK\rightarrow T_{peak}) [s]   & 383 & 43.0    &   7.58\\
        T_{peak}[10^8 K]   &   2.47    &   2.22    &   2.80\\
		T_{base} \text{ fully convective envelope }[10^7 K]   & 8.86 & 8.30 & 8.13\\
        M_{ejec}[10^{-5} M_{\odot}]    &   1.89    &   0.948    &1.17\\
        X(H)_{ejec}    &  0.286     &   0.293  &    3.12\cdot 10^{-3}\\
        X(12C)_{ejec}  &   2.61\cdot10^{-2}    &   4.63\cdot10^{-2}    &  0.326\\
        X(13C)_{ejec}  &   2.24\cdot10^{-2}       &   5.95\cdot10^{-2} &  0.515\\
        X(14N)_{ejec}  &   4.13\cdot10^{-2}    &   0.141    &  0.136\\
        X(15N)_{ejec}  &   5.64\cdot10^{-2}    &   6.51\cdot10^{-2}    &   3\cdot 10^{-3}\\
        X(16O)_{ejec}  &   5.99\cdot10^{-2}    &   0.198       &  - \\
        X(20Ne)_{ejec} &   0.176       &   1.18\cdot10^{-3}    &  - \\
        V_{ejec} (\text{mass-averaged})[km/s]    &   2666    &   3057    &3279\\
        \text{Max } V_{ejec} \text{ when }R=10^{12} cm [km/s] &   4835    &   5632    &6201\\
        K_{ejec} [10^{45} ergs]   &   1.63    &   1.03    &1.50\\
        t (T_{peak}\text{ to }R=10^{10} cm)[s]  &   371    &   319   &974\\
		\noalign{\smallskip}
		\hline
	\end{array}
	$$ 
	\caption{Results for the $1.25$ $M_{\odot}$ models}
	\label{outputShiva2}
\end{table} 
\twocolumn

\begin{table}
	$$ 
	\begin{array}{ccc}
		\hline
		\noalign{\smallskip}
		M (M_{\odot}) & \text{Model} & X(\text{7Li}) \text{ ejecta} \\
		\noalign{\smallskip}
		\hline
		\noalign{\smallskip}
		0.8 M_{\odot}  & \text{080COSHIVA}  &  1.66\cdot 10^{-6} \\
        0.8 M_{\odot}  & \text{080COcarb}  &  9.1\cdot 10^{-9} \\
        1.25 M_{\odot} & \text{125ONeSHIVA} &  1.09\cdot 10^{-6} \\
        1.25 M_{\odot} & \text{125COSHIVA}  &  1.39\cdot 10^{-5}\\
        1.25 M_{\odot} & \text{125COcarb} & 4.3\cdot 10^{-7}\\
        1.25 M_{\odot} & \text{125COSHIVA123} & 1.16\cdot 10^{-8}\\
		\noalign{\smallskip}
		\hline
	\end{array}
	$$ 
	\caption{X(7Li) ejecta in each models}
	\label{Litium}
\end{table}

\subsection{Additional consequences of the TNR}

We note that a barotropic relation has the advantage of focusing on the fundamental instability without introducing the timescales connected with energy transport, but it excludes an important feedback mechanism due to baroclinicity.  Adopting  $P(\rho)$ means that isobars are the same as constant density surfaces and, until convection begins, there is no internal source for the vorticity. The temperature fluctuations induce a $\nabla\rho\times\nabla P$ term that will produce further coupling and filamentation of the main vortices. It drives mixing but since it scales as $l^{-2}$, as does the viscosity term, the effect is on much smaller scales than the thickness of the layer. Coupling the shear with convection, driven by the TNR, should enhance the effect but such simulations require a shearing box with the nuclear energy generation and this has not yet been done (although, for a start, see \citealt{Jose2020})\footnote{The other effect of this term is that of a battery; strong (thermally induced) baroclinicity will generate a turbulent electromotive force in the dynamo equations (\citealt{Moffatt1973}, \citealt{Krause1980}) and could lead to a fast dynamo when coupled with the convection.}.

\subsection{Implications for post-outburst evolution} 

Mass transfer does not simply turn off when the nova event occurs, but the conditions that determine how the mass is accreted and stored in the circumstellar neighborhood will be very different than the pre-outburst state of the WD. Although observations point to the WD as a distended envelope far from mechanical and thermal equilibrium (e.g., \citealt{Mason2021}). This stage of recovery has still to be simulated (\citealt{Hillman}). Therefore, we suggest that shear-induced mixing could be even more important in the recovery stage than in the run-up to ignition of the TNR and could alter conclusions based on the duration of the nuclear burning. Hydrogen enrichment of the WD envelope will continue as long as it is driven within the boundary layer, and continued fueling is possible if the burning zone is sufficiently near the surface. 
This has important implications for the debate regarding the ultimate fate of the mass-gaining WD. If the burning is only because of residual fuel remaining after the envelope ejection, then the WD mass is increasing and may reach the stability limit. This is the accretion-induced collapse picture for a single degenerate origin of SN Ia. If, instead, the envelope has been completely removed or even more than the accreted mass is ejected, then this scenario cannot work. But if the temperature of the post-explosion WD envelope remains high enough, and the boundary layer pre-mixing transports hydrogen-rich but polluted gas to a sufficient depth, the nuclear burning may continue. 

What effect this will have on the subsequent explosion and ejection is, therefore, an essential application of the model discussed in this work for the boundary layer. The thermal timescale for relaxation of a WD envelope after mass ejection is from years to decades (e.g. \citealt{Mason2021}) so such short timescales for triggering a TNR require a very different treatment of the accretion process. If the remaining envelope is still hot after the ejection, and already relaxed when the accretion re-establishes, the mixing with newly accreted gas will be different. The boundary layer structure will be shearing and mixing fuel that may ignite and possibly lead to an explosive ejection of the layer without passing through the full cycle of the TNR.  It is unclear at this juncture whether the mass of the white dwarf will continue to increase or, perhaps, peg at a maximum value or even whittle down over time.

\begin{acknowledgements}
This work would not have been possible without the kind allocation of 128 cores of the {\it Cluster di calcolo ad alte prestazioni DELL POWER EDGE R630i} of San Piero a Grado
This research was supported in part by high-performance computing resources available at the Kiel University Computing Centre. 
JJ acknowledges partial support from the Spanish MINECO grant  PID2020-117252GB-I00, by the E.U. FEDER funds, and by the AGAUR/Generalitat de Catalunya grant SGR-386/2021. This article benefited from discussions within the EU’ H2020 project No. 101008324 ChETEC-INFRA. MB thanks Gianluigi Bodo and M.Cemeljic' for their precious help with PLUTO. We thank Walter Del Pozzo, Wolfgang Duschl, Ken Gayley, Ami Glasner, Tobias Illenseer, Paul Kuin, Elena Mason, and Kim Page for valuable discussions.  
\end{acknowledgements}

\begin{appendix}
\onecolumn
\section{Comparative time development of  outburst models}

\begin{figure}[h]
\centering
	{
    \begin{subfigure}{0.44\textwidth}
        \includegraphics[width=\hsize]{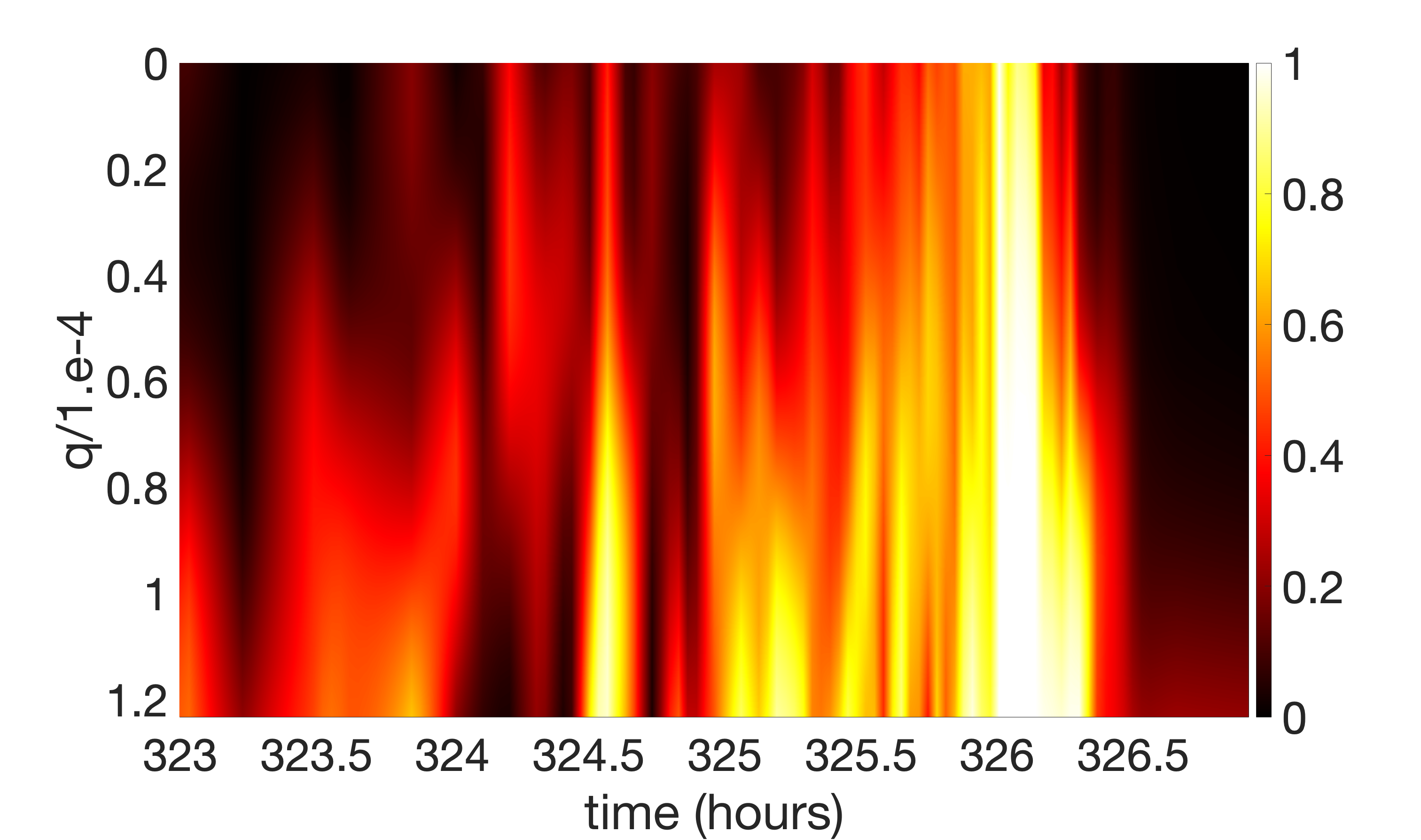}
		\caption{}
        \label{fig:conv08}
	\end{subfigure}
    \begin{subfigure}{0.44\textwidth}
        \includegraphics[width=\hsize]{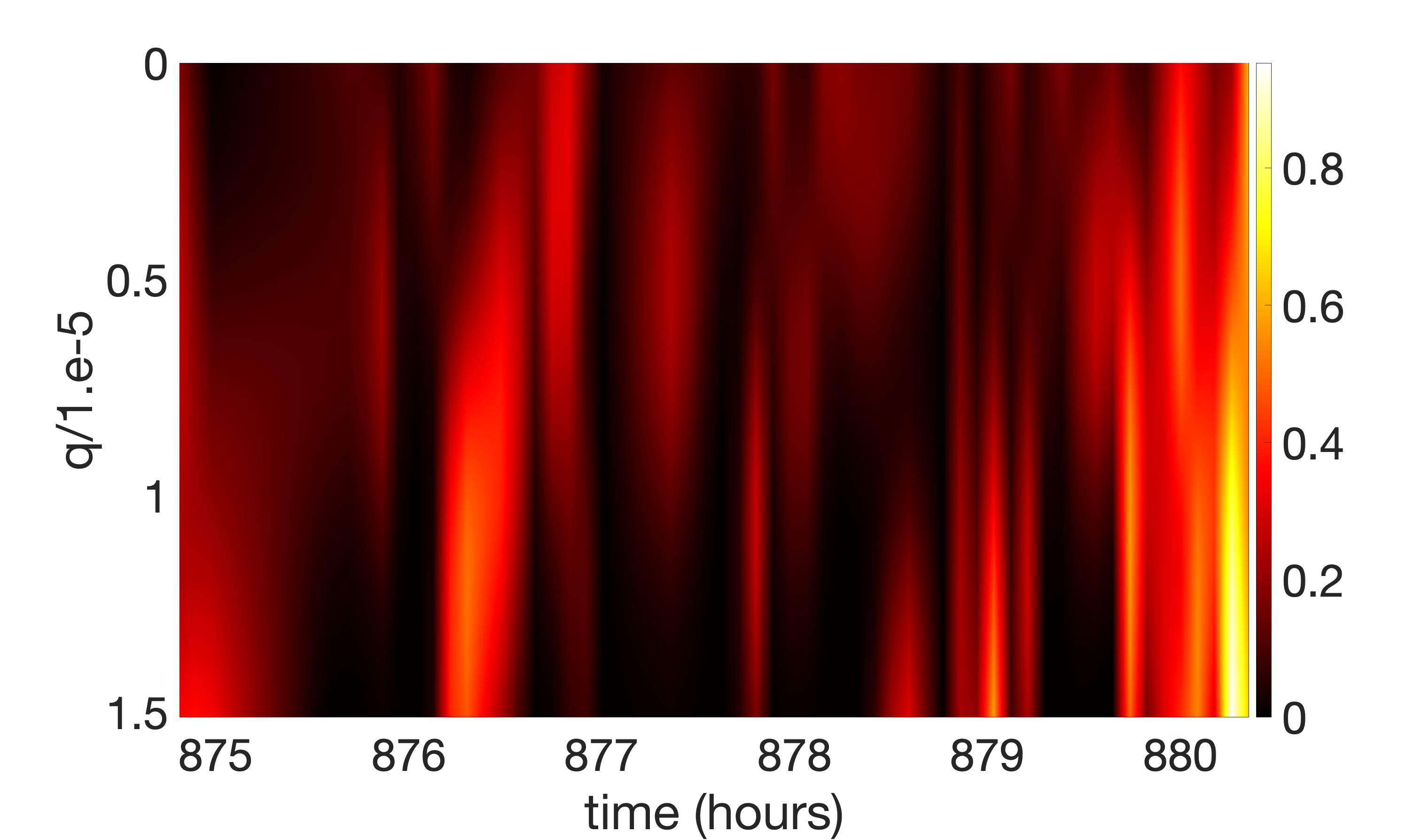}
		\caption{}
        \label{fig:conv125}
	\end{subfigure}\newline
    \begin{subfigure}{0.44\textwidth}
        \includegraphics[width=\hsize]{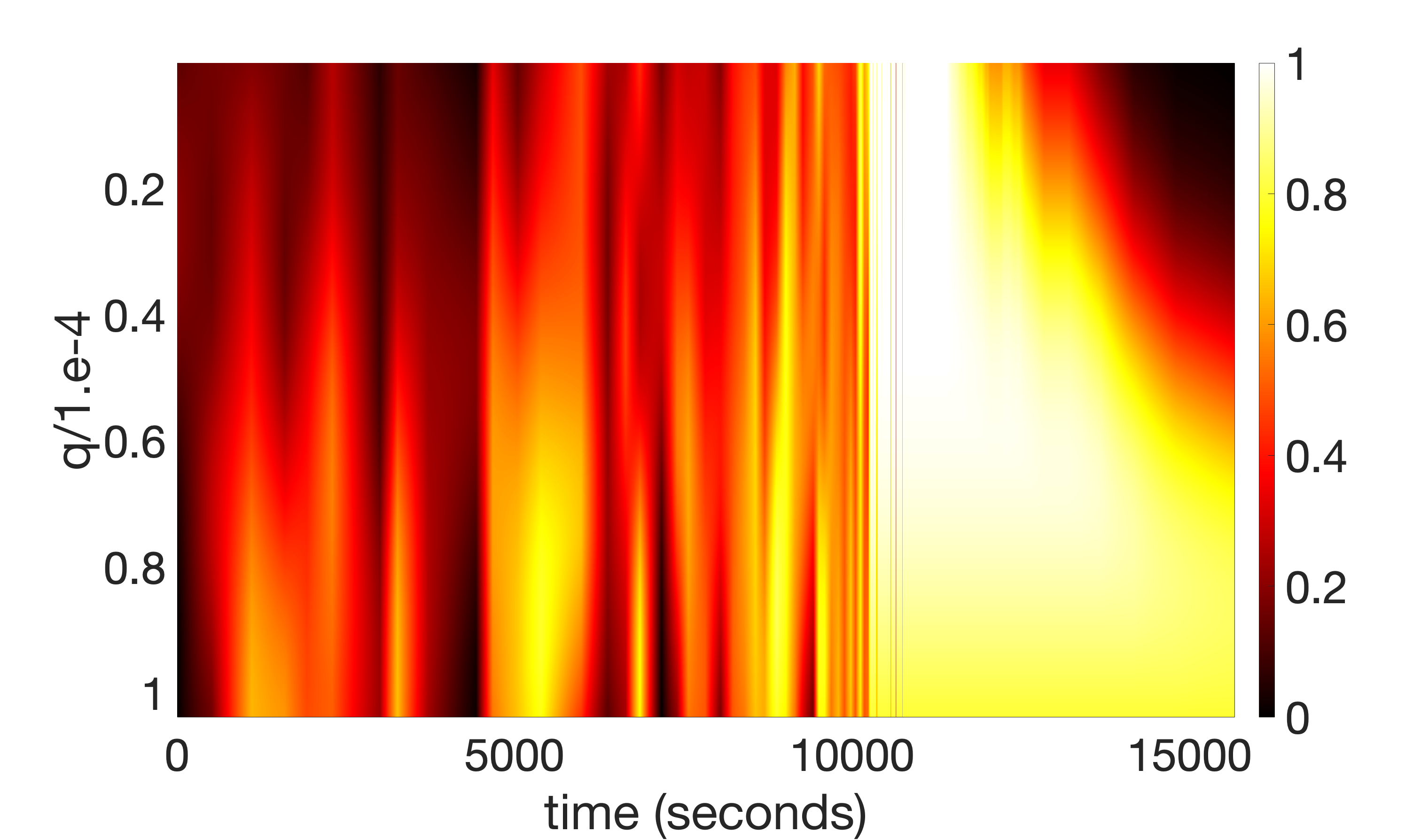}
		\caption{}
        \label{fig:conv08S}
	\end{subfigure}
    \begin{subfigure}{0.44\textwidth}
        \includegraphics[width=\hsize]{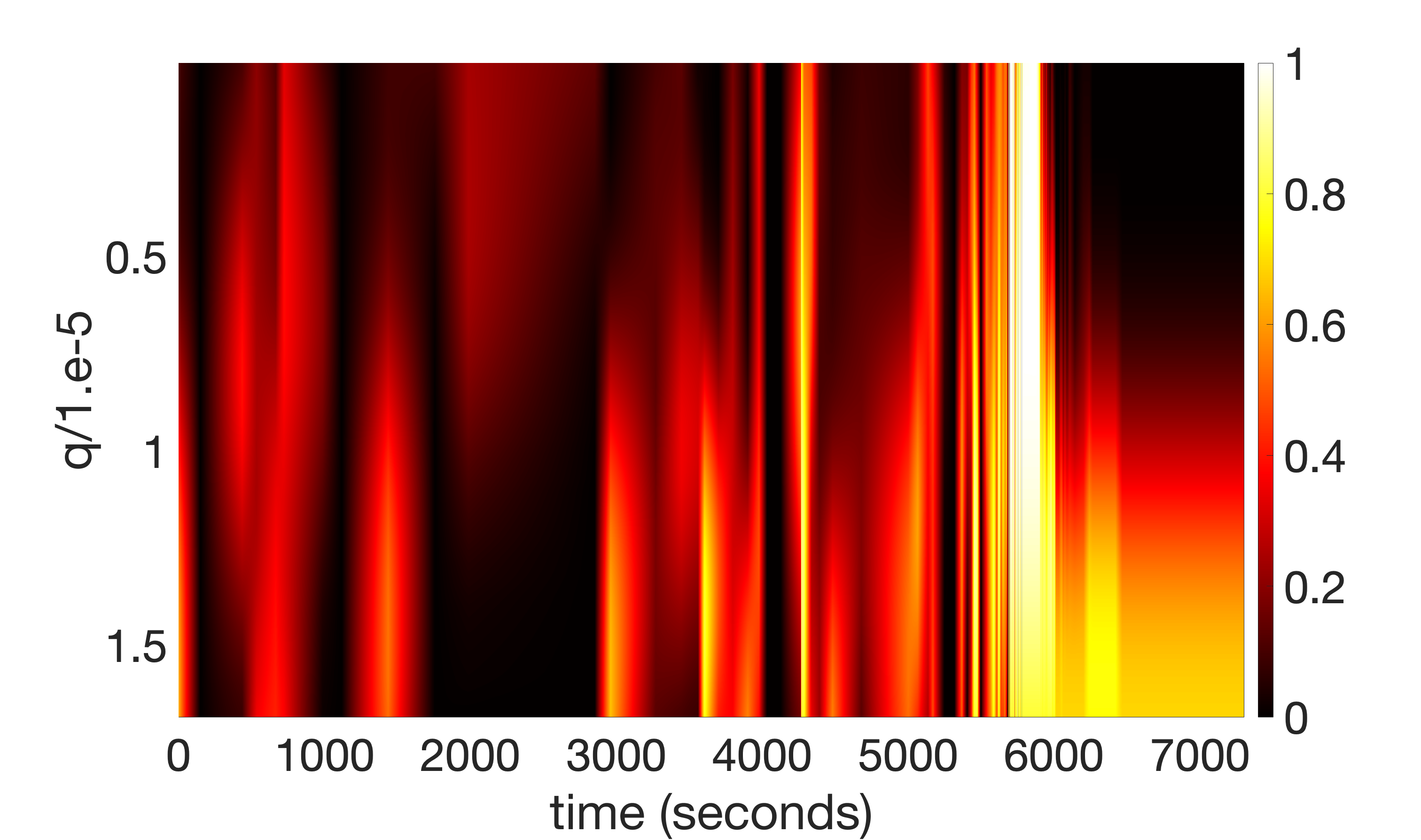}
		\caption{}
        \label{fig:conv125S}
	\end{subfigure}\newline
    \begin{subfigure}{0.44\textwidth}
        \includegraphics[width=\hsize]{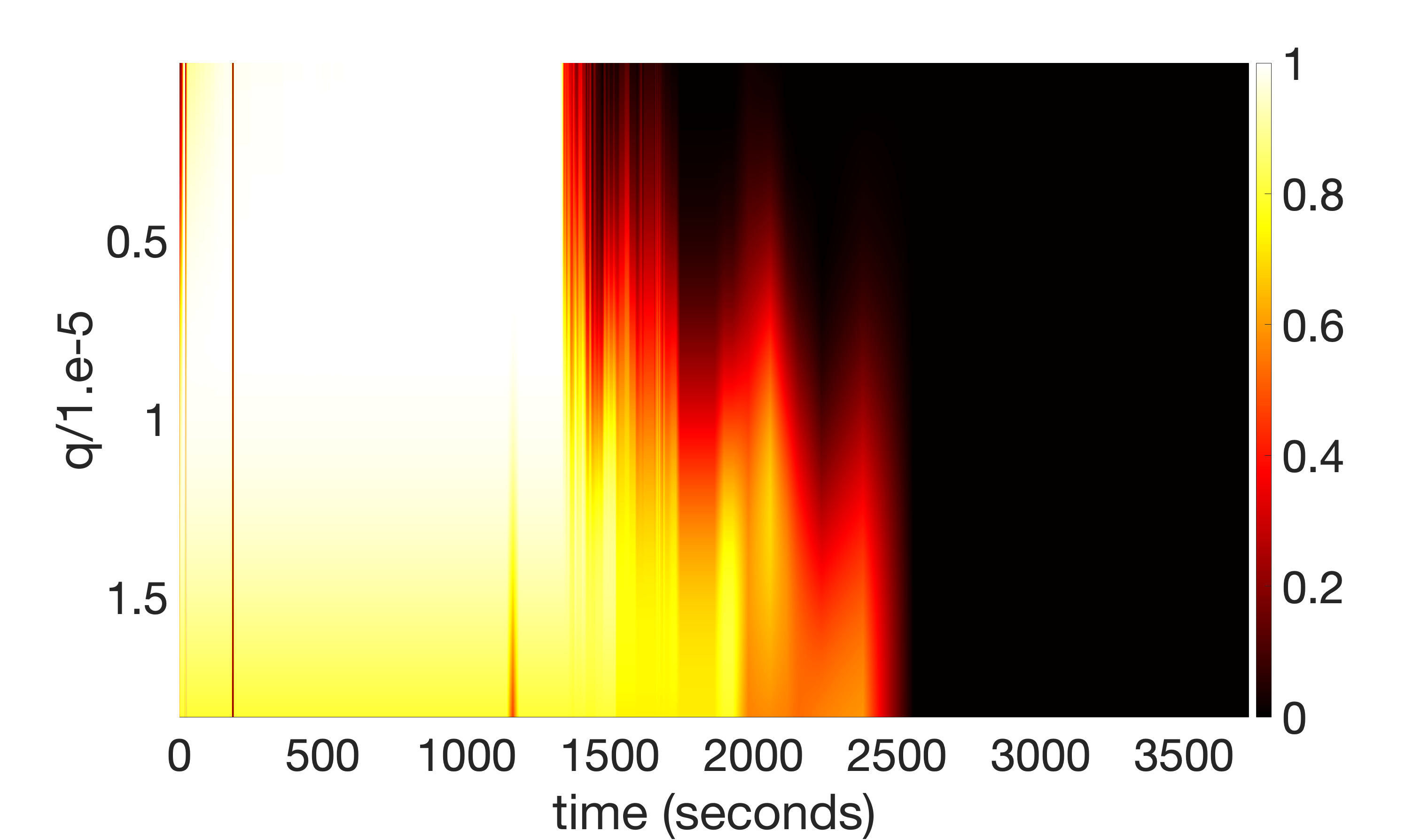}
		\caption{}
        \label{fig:conv125S123}
	\end{subfigure}
	}
	\caption{Time evolution of convection, shown as $L_{conv}/L_{tot}$, for models 080COcarb (Panel a), 125COcarb (b), 08COSHIVA (c), 125COSHIVA (d) and 125COSHIVA123 (e). On the $x$ axis there is the time, on the $y$ axis there is the mass variable q, defined in the Fig. \ref{fig:rhoev}, and the $L_{conv}/L_{tot}$ is given from the color bar. It’s important to underline that the 0 time is not the same for all the images, but these are just a zoom in the moments before the TNR.}
	\label{convcomp}	
\end{figure}

\begin{figure}[h]
\centering
	{
    \begin{subfigure}{0.44\textwidth}
        \includegraphics[width=\hsize]{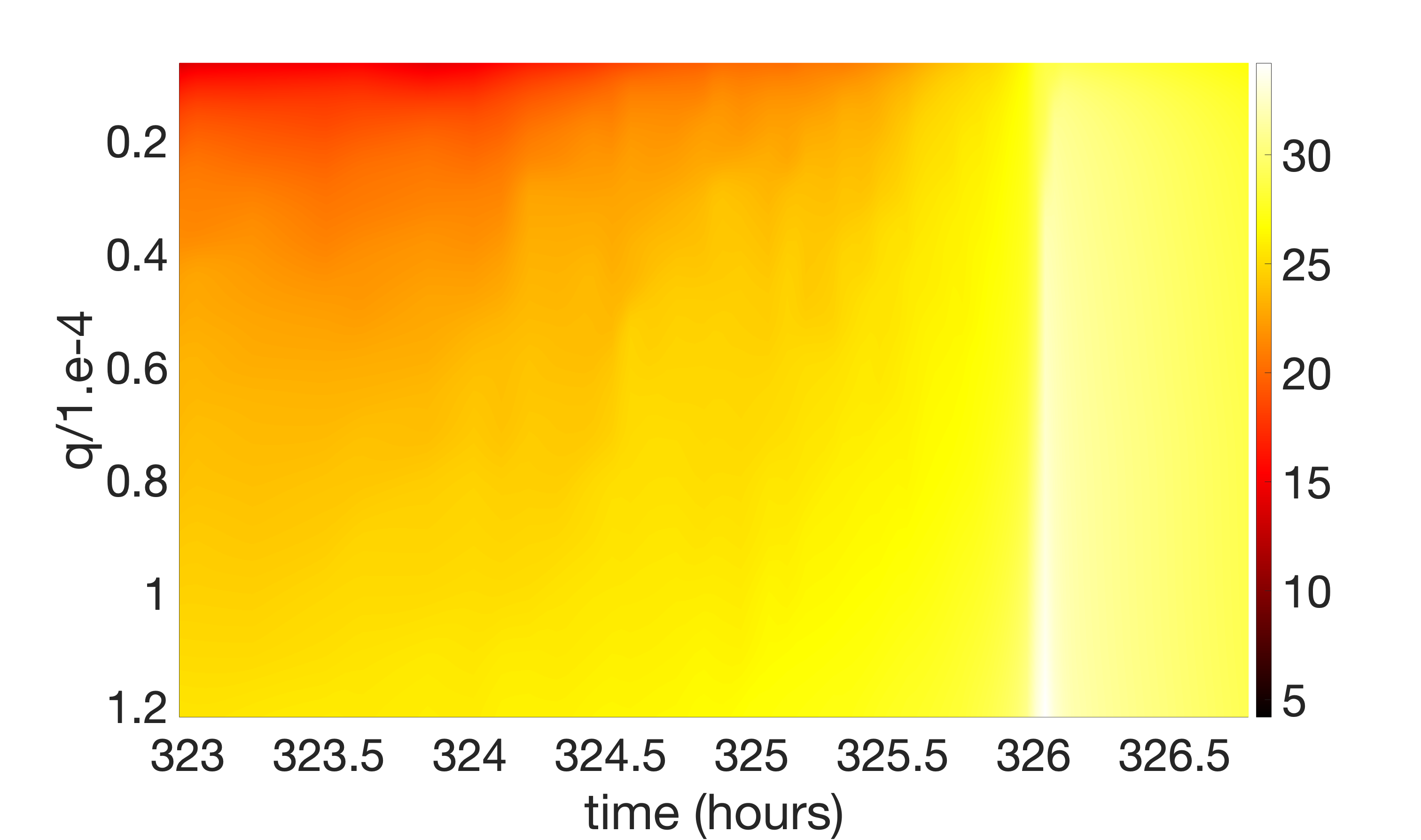}
		\caption{}
        \label{fig:enucl08}
	\end{subfigure}
    \begin{subfigure}{0.44\textwidth}
        \includegraphics[width=\hsize]{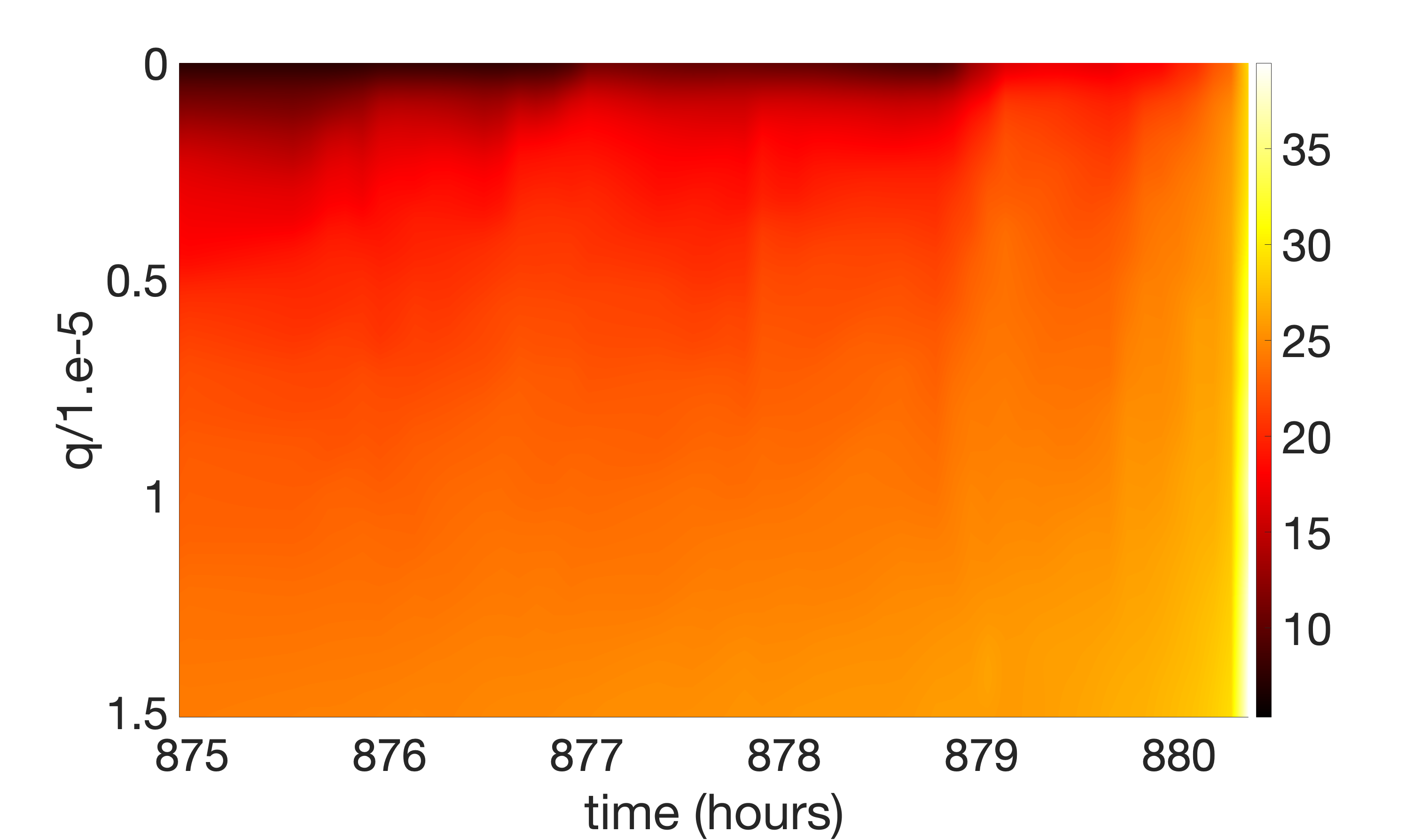}
		\caption{}
        \label{fig:enucl125}
	\end{subfigure}\newline
    \begin{subfigure}{0.44\textwidth}
        \includegraphics[width=\hsize]{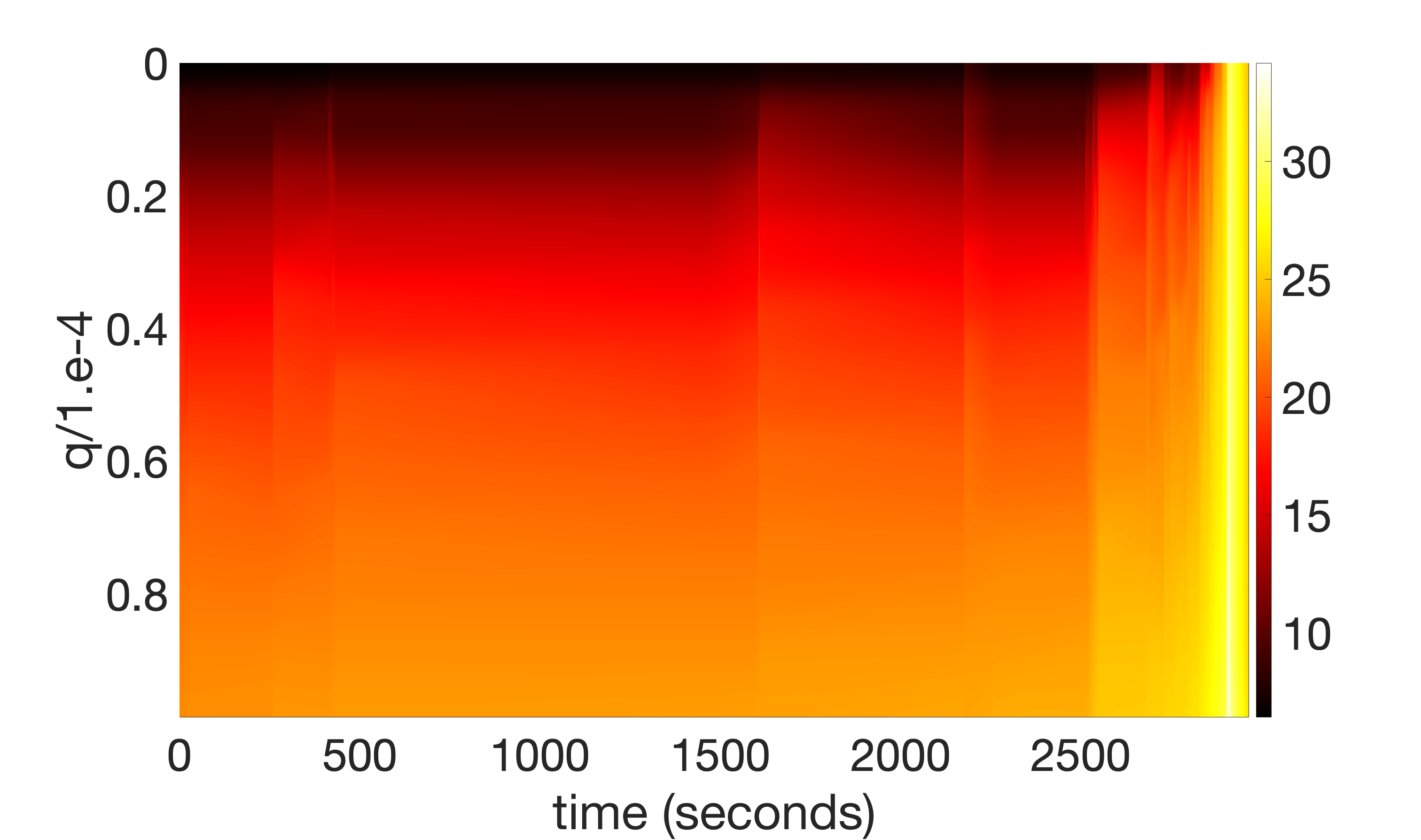}
		\caption{}
        \label{fig:Enucl08S}
	\end{subfigure}
    \begin{subfigure}{0.44\textwidth}
        \includegraphics[width=\hsize]{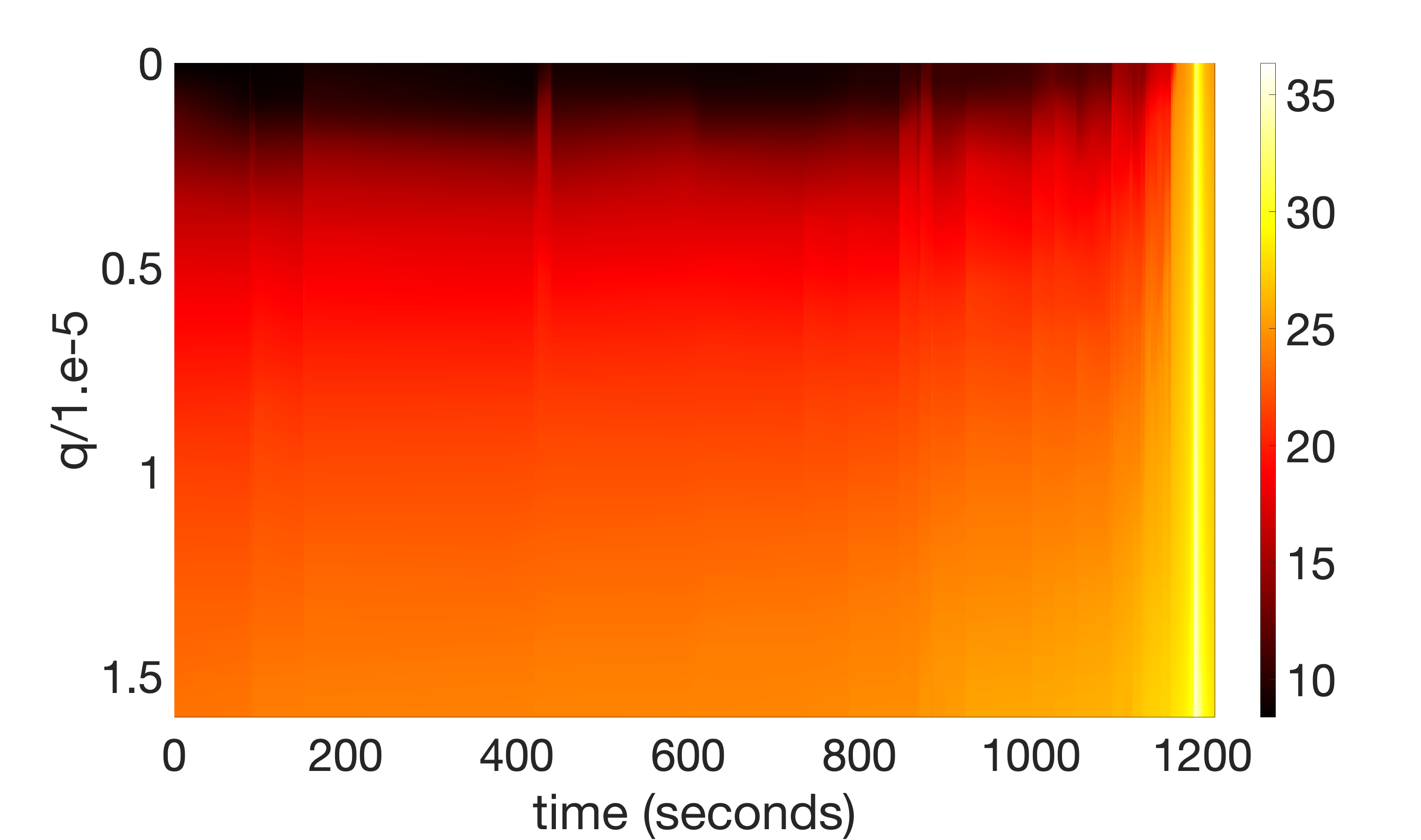}
		\caption{}
        \label{fig:Enucl125S}
	\end{subfigure}\newline
    \begin{subfigure}{0.44\textwidth}
        \includegraphics[width=\hsize]{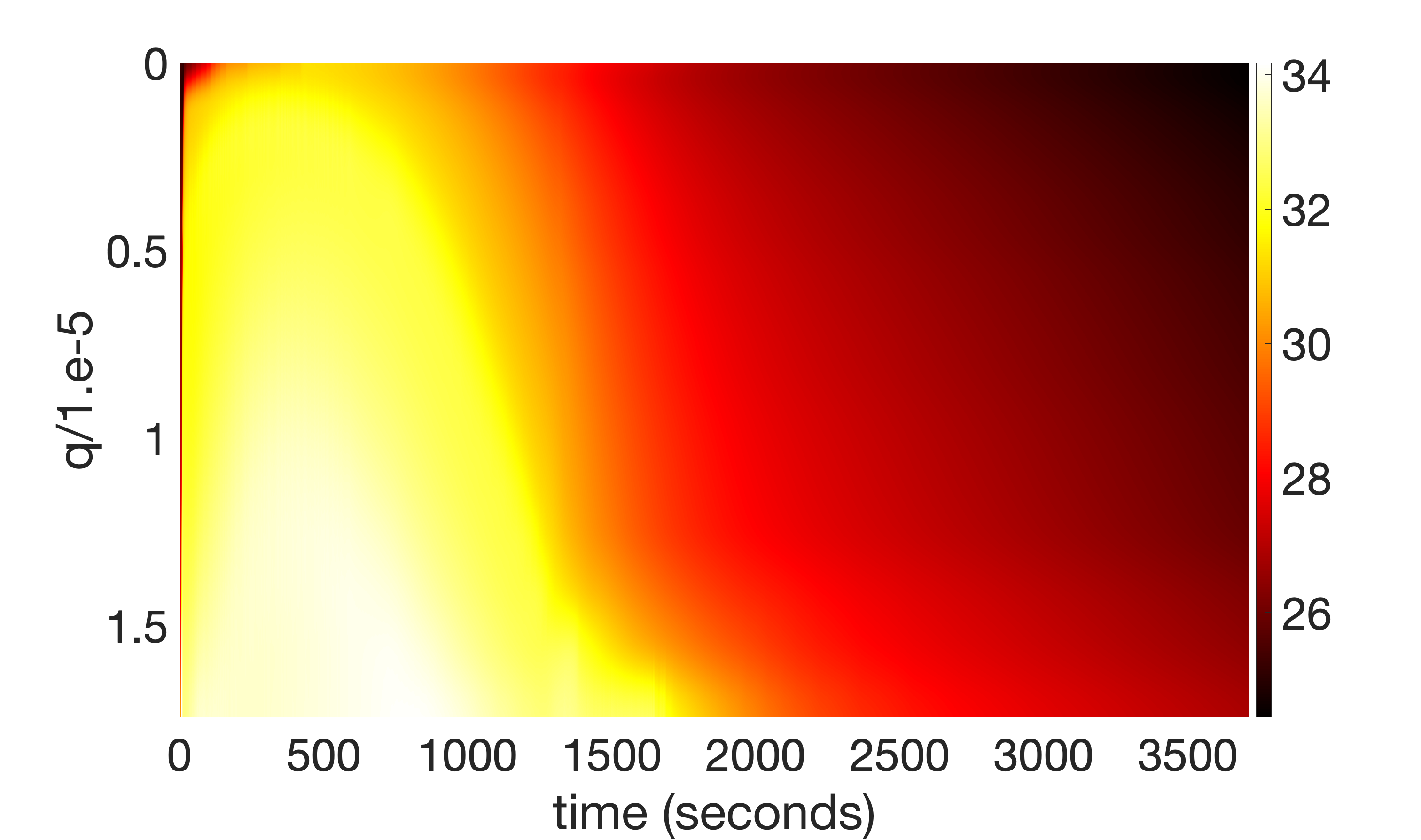}
		\caption{}
        \label{fig:Enucl125S123}
	\end{subfigure}
	}
	\caption{Same as Fig. \ref{convcomp}, but for the time evolution of the nuclear energy released, $log(E_{nuc})$.}
	\label{Enuclcomp}	
\end{figure}

In Figs. \ref{convcomp}-\ref{Enuclcomp} we show the time evolution of the convection and nuclear energy for all the CO models of the Tables \ref{outputSHIVA}-\ref{outputShiva2}.
The shear mixing profile has the same mean enrichment but because the extremes are either more hydrogen or carbon enriched, the more extended layer ignites with a greater fuel supply and thermal conduction within the layer governs the burning rate. In the standard multi-dimensional models that were the basis of the 12321 model, an arbitrary, but small, temperature perturbation was inserted at the interface as the initial condition. In the standard SHIVA models, no perturbation is needed and the accreted layer is homogenized to a fixed mixture.  If, instead, the fuel is pre-mixed with the PLUTO profile, random roundoff fluctuations suffice to instigate burning within the layer, that subsequently spreads across the zone by conduction (e.g. \citealt{Waldman2011}). The fingering observed at the onset of the 3D burning simulations  (e.g., \citealt{Casanova2016}), grows slowly since buoyancy results from the superadiabatic temperature gradient that now extends throughout the mixed layers.  The rapid burning in the most enriched portions quickly enhances the convection and the temperature rise is far more rapid than in the homogenized case.
The advance of the burning front in the standard case is initially low because the source is weak and only drives vigorous convection after a few thousand seconds (see Figs. \ref{fig:conv08S}, \ref{fig:conv125S}) about $10^{4}$ seconds for the 0.8 M$_\odot$ model and 6000 seconds for the 1.25 M$_\odot$ model). The pre-mixed layer will, in contrast, be more luminous since the burning zone is now the width of the critical layer. The reactions proceed faster and the advance of the burning front is more rapid (order of the last 1.5 hours before the TNR in Figs. \ref{fig:conv08}, and the last 30 mins before the TNR in \ref{fig:conv125}).
The convective onset is sudden and nearly complete within the conduction timescale (fig. \ref{fig:conv125S123}), in contrast to the standard models (\citealt{Jordi1998}, \citealt{Jordi2016}, \citealt{Starrfield1972}). Including the 12321 turbulent convective scheme (\citealt{Jose2020}) enhances the efficiency once buoyancy dominates the energy transfer.

\onecolumn
\end{appendix}
\end{document}